\documentclass[twocolumn,letter]{jpsj2}
\def\i{{\rm i}}
\def\d{{\rm d}}
\def\e{{\rm e}}
\def\vector#1{{\bf #1}}

\def\vp{{\vector p}}

\def\vr{{\vector r}}
\def\vR{{\vector R}}
\def\vS{{\vector S}}

\def\Tc{{T_{\rm c}}}

\def\hightc{{high-$T_{\rm c}$ }}

\def\SrRuO{\mbox{${\rm Sr_2RuO_4}$}}

\def\hsp#1{\hspace{#1ex}}

\def\Tc{{T_{\rm c}}}

\def\ppara{{p_{\parallel}}}

\def\omegaD{{\omega_{\rm D}}}

\def\eq.#1{eq.~(\ref{#1})} 

\newcommand\Equation[3]{
\begin{equation}\label{#1}\tag{#2}
#3
\end{equation}
}

\title{
Internal Transition of Superconducting State by Impurity Doping \\ 
with a Jump of Isotope-effect Coefficient in Multiband \\ 
Superconductors 
}

\author{Hiroshi {\sc Shimahara}}

\def\runauthor{Hiroshi {\sc Shimahara}}

\inst{
Department of Quantum Matter Science, ADSM, Hiroshima University, 
Higashi-Hiroshima 739-8530, Japan
}

\recdate{April ~~, 2003}

\abst{
The impurity effects on the transition temperature $\Tc$ and 
the isotope effect are examined in multiband superconductors 
with magnetic and nonmagnetic impurities, 
where the effect of Coulomb repulsion is considered. 
It is shown that an internal transition of the superconducting state 
is induced by impurity doping, 
and that the transition is accompanied by a jump of 
the isotope-effect coefficient $\alpha$. 
In particular, the transition is illustrated in a system 
with two electron bands. 
In some special cases, 
extended Abrikosov and Gor'kov (AG) equations for $\Tc$ 
and the expressions of the isotope-effect coefficient $\alpha$ 
are obtained. 
Possible relevance of the present mechanism to the experimental results 
of \SrRuO \hsp{0.25} is discussed. 
}

\kword{
Abrikosov-Gor'kov equation, 
superconducting transition temperature, 
isotope effect, 
multiband superconductor, impurity effect, 
weakly screened electron-phonon interaction 
}

\begin{document}
\sloppy
\maketitle



The isotope-effect coefficient provides important information on 
the mechanism of superconductivity. 
For example, 
the BCS theory based on the electron-phonon interactions 
is supported in nontransition-metal superconductors, 
such as ${\rm Hg}$ and ${\rm Zn}$, 
by their isotope-effect coefficient $\alpha$, 
which is nearly equal to 0.5. 
In transition metals and some compounds, large deviations 
from $\alpha = 0.5$ have been observed, 
but the deviations can be explained in the context of electron-phonon 
interactions by taking into account strong Coulomb repulsion, 
anharmonicity of lattice vibrations, 
and van Hove singularity~\cite{Note01}.

On the other hand, in exotic superconductors, such as 
\hightc cuprates, organic superconductors, 
\SrRuO, and heavy fermion superconductors, 
nonphonon mechanisms of superconductivity, 
such as spin-fluctuation exchange interactions, 
have been examined as possible mechanisms. 
In such nonphonon mechanisms, 
the isotope-effect coefficient must deviate from 0.5 markedly, 
or vanish completely. 
The presence of the isotope shift ($\alpha \ne 0$) suggests 
the presence of a contribution to the pairing interaction 
from electron-phonon interactions.

Therefore, it may be difficult to derive information 
on the pairing mechanism solely from a single value of $\alpha$. 
Therefore, 
it is useful to systematically examine correlations between $\alpha$ and 
other quantities, such as impurity concentration, transition temperature, 
hole concentration (for \hightc cuprates), and pressure.

We examine the impurity effect on the isotope effect in this study. 
First, we briefly review it in single-band systems. 
In the presence of impurities, $\Tc$ is reduced except 
for nonmagnetic isotropic impurity scattering and isotropic pairing. 
The ratio $\Tc/\Tc_0$ satisfies 
the Abrikosov and Gor'kov (AG) equation~\cite{Abr61} 
\Equation{eq:AGformula}
{1}{
     \ln \bigl ( \frac{\Tc_0}{\Tc} \bigr ) 
          =   \psi \bigl ( 
                      \frac{1}{2} + \frac{\zeta_{\rm s}}{2\pi \Tc}
                   \bigr ) 
            - \psi \bigl ( \frac{1}{2} \bigr ) 
          \equiv \Phi(\zeta_{\rm s}) , 
     }
where $\zeta_{\rm s}$ is a scattering rate proportional 
to $n_{\rm imp}$, and $\psi(x)$ denotes the digamma function. 
We have introduced a function $\Phi$ for convenience.

If a slight change in the atomic mass does not affect the mechanisms of 
the pairing interactions and the pair-breaking effect by impurities, 
it is plausible to assume that the same equation as \eq.{eq:AGformula} 
\Equation{eq:AGformulatwo}
{2}{
     \ln \bigl (
          \frac{\Tc(M',0)}{\Tc(M',\zeta_{\rm s})}
         \bigr ) 
          =   \psi \bigl (  
                   \frac{1}{2} 
                   + \frac{\zeta_{\rm s}}{2 \pi \Tc(M',\zeta_{\rm s})}
                   \bigr ) 
            - \psi \bigl ( \frac{1}{2} \bigr ) 
     }
holds in the system with an atomic mass $M'$ 
slightly different from $M$. 
From eqs.~(\ref{eq:AGformula}) and (\ref{eq:AGformulatwo}), 
we immediately obtain 
\Equation{eq:isotopeuniversal}
{3}{
     \frac{\alpha}{\alpha_0}
     = \Bigl [  1  - 
          \frac{\zeta_{\rm s}}{2\pi \Tc} 
          \psi' \bigl ( \frac{1}{2} + \frac{\zeta_{\rm s}}{2\pi \Tc}
                \bigr )
       \Bigr ]^{-1} , 
     }
which has been obtained by Carbotte, Greeson and Perez-Gonzalez 
(CGP)~\cite{Car91}, where 
${
     \alpha = - {\partial \ln \Tc}/{\partial \ln M} 
     }$ 
and $\alpha_0 = \alpha(M,0)$. 
It should be noted that 
the relation between $\alpha/\alpha_0$ and $\Tc/T_{{\rm c}0}$, 
which is obtained by eliminating $\zeta_{\rm s}$ 
from eqs.~(\ref{eq:AGformula}) and (\ref{eq:isotopeuniversal}), 
is universal in the sense that it does not depend on 
the mechanism of the pairing interactions, 
properties of impurity scattering, 
and even the pairing symmetry.

In \SrRuO, however, a large deviation from the CGP universal relation 
has been observed by Mao {\it et al.}\cite{Mao01} 
When the transition temperature is lowered by defects (as impurities), 
the coefficient $\alpha$ rapidly increases 
near $\Tc \approx 0.94 T_{{\rm c}0}$ 
from the negative value $\alpha \approx - 0.1$ near $\Tc = \Tc_{0}$ 
to the positive value $\alpha \approx 0.2$~\cite{Mao01}. 
Below $\Tc \approx 0.94 T_{{\rm c}0}$, 
$\alpha$ seems to follow the CGP relation. 
On the other hand, it has been experimentally suggested that 
$\Tc$ obeys \eq.{eq:AGformula}~\cite{Mao99}. 
Since the CGP relation is derived directly 
from \eq.{eq:AGformula} as explained above, 
the large deviation from it appears mysterious.

In this paper, we develop a formulation of the impurity effect 
on $\Tc$ and $\alpha$ in multiband superconductors, 
and show that an abrupt change in the isotope coefficient $\alpha$ 
with impurity doping is possible in some conditions, 
even when $\Tc$ practically obeys an AG equation.

The impurity effect in the multiband superconductor has been examined by 
some authors~\cite{Gol97,Ars03,Agt99}. 
Golubov and Mazin examined this subject and 
calculated the density of states~\cite{Gol97}. 
Agterberg examined the impurity effect in a two-band case 
for \SrRuO \hsp{0.25} in the context of the orbital-dependent 
superconductivity~\cite{Agt99,Agt97}. 
It is known that \SrRuO \hsp{0.25} is 
a multiband superconductor~\cite{Ogu95,Sin95}.

The appreciable isotope shifts observed in \SrRuO~\cite{Mao01} 
suggest the presence of the phonon contribution.~\cite{Shi03} 
For simplicity, we ignore any possible additional nonphonon pairing 
interactions~\cite{Note04}, 
but the present theory is applicable 
whether they exist or not~\cite{Note02}. 
It should be noted that anisotropic pairing can be induced 
by the phonon interactions~\cite{Fou77,Shi02a,Shi02b}.

We define the impurity potential which scatters electrons 
from $X$-band to $X'$-band by 
\Equation{eq:impuritypotentialdef}
{4}{
     u_{XX'}(\vr - \vR_i) 
     + v_{XX'}(\vr - \vR_i) \, 
          \vS_i \cdot {\mib \sigma}(\vr) . 
     }
We introduce symmetry functions $\gamma^{(X)}_{\alpha}(\ppara)$ 
with index $\alpha$ by which the pairing interaction $V_{XX'}(\vp,\vp')$ 
is written in a diagonal form as 
\Equation{eq:pairinginteractionexpanded}
{5}{
     V_{XX'}(\vp, \vp') 
          = - \sum_{\alpha} \gamma^{(X)}_{\alpha}(\ppara) \, 
               g^{(\alpha)}_{XX'} 
               \, \gamma^{(X')}_{\alpha}(\ppara') . 
     }
Here, $\ppara$ as an argument of $\gamma^{(X)}_{\alpha}$ denotes 
the two-dimensional momentum coordinate on the Fermi surface of the $X$-band, 
where band suffixes are omitted from it. 
For example, in the isotropic case, $\ppara$ can be replaced 
with the polar coordinates $(\theta,\varphi)$, 
and we can put $\alpha = s, p_x, p_y, \cdots$. 
The symmetry functions $\gamma^{(X)}_{\alpha}$ are defined so that 
the orthonormal condition 
${
    \int \d^2 \ppara \rho_X(0,\ppara) 
         \gamma^{(X)}_{\alpha} (\ppara) \gamma^{(X)}_{\alpha'} (\ppara) 
         /  N_X(0) 
    = \delta_{\alpha \alpha'} 
    }$ 
is satisfied, 
where $N_X(0)$ and $\rho_X(0,\ppara) \d^2 \ppara$ denote 
the density of states of the $X$-band at the Fermi energy and 
that within the infinitesimal area $\d^2 \ppara$ near $\ppara$, 
respectively. 
In particular, we put $\gamma^{(X)}_{s} = 1$.

The gap function in the $X$-band is expanded by the symmetry functions as 
${
     \Delta_X(\vp) = 
          \sum_{\alpha} \Delta_{X \alpha} 
               \gamma^{(X)}_{\alpha}(\ppara) 
     }$. 
The impurity potentials are also expanded as 
\Equation{eq:impuitypotentialexpanded}
{6}{
     |u_{XX'}(\vp, \vp')|^2 
          = \sum_{\alpha} \gamma^{(X)}_{\alpha}(\ppara) \, 
               u^{(2\alpha)}_{XX'} \, \gamma^{(X')}_{\alpha}(\ppara') 
     }
and the same for $|v_{XX'}(\vp, \vp')|^2$, 
which define $u^{(2\alpha)}_{XX'}$ and $v^{(2\alpha)}_{XX'}$, 
where the pairing interactions off-diagonal with respect to the suffix 
$\alpha$ are ignored for simplicity.

Here, we neglect the momentum dependences of 
$g_{XX'}^{(\alpha)}$, $u$ and $v$, in the direction perpendicular to 
the Fermi surface, 
except that we introduce the cutoff energy 
$\omega_{\rm c}$ ($\sim \omegaD$) 
for the pairing interactions $g_{XX'}^{(\alpha)}$.

For an ``$\alpha$''-wave superconductor, the gap equation is obtained 
by the Born approximation~\cite{Gol97} as 
\Equation{eq:gapeqinBornappr}
{7}{
     \Delta_{X \alpha} 
          = 2 \pi T \sum_{X'} \sum_{n'= 0}^{n_{\rm c}} 
            \lambda_{XX'}^{(\alpha)}
            \frac{\tilde \Delta_{X'n' \alpha}}
                 {{\tilde \omega}^{(X')}_{n'} } , 
     }
with 
\Equation{eq:Deltatildedef}
{8}{
     \begin{split}
     {\tilde \Delta}_{Xn \alpha} 
          & \equiv 
               \Delta_{X \alpha} + \sum_{Y} M_{XY} 
               {\tilde \Delta}_{Yn \alpha} \\[-4pt]
     \i {\tilde \omega}^{(X)}_{n} 
          & \equiv 
               \i \omega_n + \i \, {\rm sgn} (\omega_n) / 2 \tau_{1X} . 
               \\
     \end{split}
     }
Here, we have put $T = \Tc$, 
and defined 
$\lambda_{XX'}^{(\alpha)} = g_{XX'}^{(\alpha)} N_{X'}(0)$, 
$M_{XY} = \Lambda_{XY}/|{\tilde \omega}^{(Y)}_{n}|$, 
$\Lambda_{XY} = 1/2 \tau_{2XY}$, 
$ 1/\tau_{1X} =  \sum_{Y} 1/\tau_{1XY} $, 
and 
\Equation{eq:tauexpression}
{9}{
     \begin{split}
     \frac{1}{2 \tau_{1XY}} 
          & =      \pi n_{\rm imp} N_Y(0) 
                    \Bigl [   
                        u_{XY}^{(2 s)}
                      + v_{XY}^{(2 s)} S(S + 1) 
                    \Bigr ] ,  \\
     \frac{1}{2 \tau_{2XY}} 
          & =      \pi n_{\rm imp} N_Y(0) 
                    \Bigl [   
                        u_{XY}^{(2 \alpha)} 
                      - c_{\alpha} v_{XY}^{(2 \alpha)} S(S + 1) 
                    \Bigr ]  , \\
     \end{split}
     }
where $c_{\alpha} = 1$ and $1/3$ when ``$\alpha$''-wave pairing 
is of singlet and triplet pairings, respectively. 
We introduce matrices 
${\hat M}$, ${\hat \lambda}$, and ${\hat \omega}$ 
whose $XY$ elements are $M_{XY}$, $\lambda_{XY}^{(\alpha)}$, 
and ${\tilde \omega}^{(X)}_{n'} \delta_{XY}$, respectively, 
and a vector ${\hat \Delta}$ whose $X$ element is $\Delta_{X\alpha}$. 
Then, we could rewrite \eq.{eq:gapeqinBornappr} as 
$ {\hat \Delta} =   {\hat \lambda} \, {\hat K} {\hat \Delta}$, 
with 
${
     {\hat K } \equiv 2 \pi T \sum_{n' = 0}^{n_{\rm c}} 
            {\hat \omega}^{-1} \, 
            ( 1 - {\hat M} )^{-1} 
      }$. 
Furthermore, if we define 
\Equation{eq:Lhatdef}
{10}{
     {\hat L} = {\hat \lambda}^{-1} + {\hat \Phi}
     }
with ${\hat \Phi} = L {\hat I} - {\hat K}$, 
$L = \ln (2 \e^{\gamma} \omega_{\rm c}/\pi \Tc)$, 
and the Euler constant $\gamma = 0.57721\cdots$, 
the gap equation can be rewritten in a compact form 
\Equation{eq:gapeqwithmatrix}
{11}{
     {\hat L} {\hat \Delta} = L {\hat \Delta} . 
     }
The matrix ${\hat \Phi}$ includes all information of the impurity 
scattering. 
With the smallest positive eigenvalue $L$ of the matrix ${\hat L}$, 
$\Tc$ is expressed as 
\Equation{eq:Tcgeneral}
{12}{
     \Tc = \frac{2 \e^{\gamma}}{\pi} \omega_{\rm c} \exp[ - L ] . 
     }

In particular, 
when $\Lambda_{XY} \equiv 1/2\tau_{2XY} = 0$ for $X \ne Y$, 
the $XY$ element of ${\hat \Phi}$ becomes 
$\Phi_{XY} = \Phi(\zeta_{X}') \delta_{XY}$ with 
\Equation{eq:zetaAprimeLambdaABdef}
{13}{
     \zeta'_X \equiv \frac{1}{2 \tau_{1X}} - \frac{1}{2 \tau_{2XX}} . 
     }
In the absence of the interband Cooper pair hopping 
($g_{XY} = g_{XX} \delta_{XY}$), 
we obtain 
${
     \ln (T_{{\rm c}0X}/\Tc) = \Phi(\zeta'_X) 
     }$ 
with $T_{{\rm c}0X} = (2\e^{\gamma}/{\pi})
     \omega_{\rm c} \exp[-1/\lambda_{XX}]$, 
which are the same as the AG equation 
except that $\zeta'_X$ includes $1/2 \tau_{1XY}$ with $X \ne Y$.

In the general case, {\it i.e.}, 
when $\Lambda_{XY} \equiv 1/2\tau_{2XY}$ is not necessarily 
equal to $0$ for $X \ne Y$, it is difficult to write the equation 
in a simple form such as that in the above. 
Therefore, we examine systems with two bands $A$ and $B$ 
to illustrate the mechanism of the transition. 
The eigenvalues of ${\hat L}$ are obtained as 
\Equation{eq:Lexpression}
{14}{
     L = \frac{1}{2} 
          \Bigl [
               {\rm tr} ({\hat L}) 
               \pm \sqrt{ \bigl ( {\rm tr} ({\hat L}) \bigr )^2 
                          - 4 {\rm \det} ({\hat L}) } 
          \Bigr ]  . 
     }
The smaller positive one gives $\Tc$ using \eq.{eq:Tcgeneral}. 
The explicit forms of the matrix elements of ${\hat \Phi}$ are obtained 
by carrying out the summation over $n'$ as 
\Equation{eq:Phihatelementdef}
{15}{
     \begin{split}
     \Phi_{XX} 
        & =   \frac{1}{2} \bigl ( 
                  1 + \frac{\zeta_{AB}'}{\zeta_{AB}} 
                          \bigr ) 
                \Phi(\zeta_X)
            + \frac{1}{2} \bigl ( 
                  1 - \frac{\zeta_{AB}'}{\zeta_{AB}} 
                          \bigr ) 
                \Phi(\zeta_{\bar X}) \\
     \Phi_{X {\bar X}} 
        & =   \frac{\Lambda_{X{\bar X}}}{\zeta_{AB}} 
                (\Phi(\zeta_B) - \Phi(\zeta_A)) , 
     \end{split}
     }
with ${\bar A} \equiv B$, ${\bar B} \equiv A$, 
$\zeta_{AB} = \zeta_A - \zeta_B$, and $\zeta'_{AB} = \zeta'_A - \zeta'_B$, 
where $\zeta = \zeta_A$ and $\zeta_B$ are the solutions to the equation 
${
     (\zeta - \zeta'_A) (\zeta - \zeta'_B) 
       - \Lambda_{AB} \Lambda_{BA} = 0 
     }$, 
that is, 
${
     \zeta_{A, B} = \frac{1}{2} 
          [ ( \zeta_A' + \zeta_B' ) 
            \pm \sqrt{   ( \zeta_A' - \zeta_B' )^2 
                       + 4 \Lambda_{AB} \Lambda_{BA} } \, ] 
     }$. 
Thus, the right-hand side of \eq.{eq:Lexpression} can be easily evaluated 
from \eq.{eq:Lhatdef}.

Now, we examine the Coulomb effects on the superconductivity. 
In the single-band system, it is known that 
the Coulomb interaction $U$ can be taken into account by 
replacing $\lambda$ with ${\tilde \lambda} \equiv \lambda - \mu^{*} $, 
where $\mu^{*} = {\tilde U} N(0)$ and 
${\tilde U} = U/(1 + U N(0) \ln (W/\omega_{\rm c}))$. 
We have introduced an effective cutoff energy $W$ of the Coulomb interaction, 
which is of the order of the band width. 
Since $\mu^{*}$ includes $\omega_{\rm c}$, it affects the isotope effect.

In the multiband system, it is found that 
the effective Coulomb interaction is obtained by 
\Equation{eq:Utildemultiband}
{16}{
     {\tilde U}^{(\alpha)}_{XX'} 
     = U^{(\alpha)}_{XX'} 
       - \sum_{X''} U^{(\alpha)}_{XX''} L^{X''}_{\rm C} 
                    {\tilde U}^{(\alpha)}_{X''X'} , 
     }
with $L_{\rm C}^{X''} = N_{X''}(0) \ln (W/\omega_{\rm c})$, 
when the Coulomb interaction is expanded in a similar manner to 
that in eqs.~(\ref{eq:pairinginteractionexpanded}) 
and (\ref{eq:impuitypotentialexpanded}). 
We have assumed that 
$U^{(\alpha)}_{XX'}$ and ${\tilde U}^{(\alpha)}_{XX'}$ are constants 
within the ranges $|\xi_{\vp}|$, $|\xi_{\vp'}| < W$ 
              and $\omega_{\rm c}$, respectively, 
while outside these ranges, $U^{(\alpha)}_{XX'} = 0$ and 
${\tilde U}^{(\alpha)}_{XX'} = 0$, 
where $\xi_{\vp}$ denotes the single-particle energy measured 
from the Fermi energy.

In two-band systems, we obtain 
\Equation{eq:Utildeexpression}
{17}{
     \begin{split}
     {\tilde U}^{(\alpha)}_{XX} 
        & = \frac{1}{D}
            \Bigl [
               (1 + L^{{\bar X}}_{\rm C} U^{(\alpha)}_{{\bar X}{\bar X}}) 
                                         U^{(\alpha)}_{XX} 
               - L^{X}_{\rm C} U^{(\alpha)}_{X{\bar X}} 
                               U^{(\alpha)}_{{\bar X}X} 
            \Bigr ] \\
     {\tilde U}^{(\alpha)}_{X{\bar X}} 
        & = \frac{1}{D}
            \Bigl [
               (1 + L^{X}_{\rm C} U^{(\alpha)}_{XX}) 
                                  U^{(\alpha)}_{{\bar X}X} 
               - L^{{\bar X}}_{\rm C} U^{(\alpha)}_{{\bar X}X} 
                                      U^{(\alpha)}_{XX} 
            \Bigr ] , \\
     \end{split}
     }
with 
${
     D = (1 + L^{A}_{\rm C} U_{AA}^{(\alpha)}) 
         (1 + L^{B}_{\rm C} U_{BB}^{(\alpha)}) 
            - L^{A}_{\rm C} U_{AB}^{(\alpha)} 
              L^{B}_{\rm C} U_{BB}^{(\alpha)} 
     }$. 
The coupling constants 
$\lambda^{(\alpha)}_{XX'}$ should be replaced with 
${\tilde \lambda}^{(\alpha)}_{XX'} 
     = \lambda^{(\alpha)}_{XX'} - \mu^{*}_{XX'}$, 
where $\mu^{*}_{XX'} \equiv {\tilde U}^{(\alpha)}_{XX'} N_{X'}(0)$. 
The matrix ${\hat L}$ is redefined by 
${\hat L} \equiv ({\hat \lambda} - {\hat \mu^{*}})^{-1} + {\hat \Phi}$ 
with a matrix ${\hat \mu^{*}}$ whose $XY$ element is $\mu^{*}_{XY}$.

The transition temperatures and the isotope-effect coefficients can be 
explicitly calculated from the set of equations obtained above. 
However, for simplicity and in order to clarify a physical picture, 
we concentrate ourselves on a symmetric case in which two bands are equal, 
namely, $N_A(0) = N_B(0) \equiv N(0)$, 
$f_{AA} = f_{BB}$ and $f_{AB} = f_{BA}$, 
where $f = g^{(\alpha)}, u^{(2\alpha)}, v^{(2\alpha)}$ and $U^{(\alpha)}$. 
In this case, the gap functions ${\hat \Delta}^{(\pm)}$ with 
$\Delta_{A\alpha} = \pm \Delta_{B\alpha}$ are the solutions of 
\eq.{eq:gapeqwithmatrix}. 
The eigenvalues are 
\Equation{eq:Lplusminus}
{18}{
     L_{(\pm)} = \frac{1}{{\tilde \lambda}_{(\pm)}}
                 + \Phi ( \zeta^{(\pm)} ) , 
     }
where 
${
     {\tilde \lambda}_{(\pm)} 
     \equiv 
     {\tilde \lambda}_{AA} \pm {\tilde \lambda}_{AB}
     = 
     \lambda_{(\pm)} - \mu^{*}_{(\pm)}
     }$
with $\lambda_{(\pm)} \equiv \lambda_{AA} \pm \lambda_{AB}$ 
and $\mu^{*}_{(\pm)} \equiv \mu^{*}_{AA} \pm \mu^{*}_{AB}$, and 
${
     \zeta^{(\pm)} = \zeta' \mp \Lambda_{AB} 
     }$ 
with $\zeta' \equiv \zeta'_{A} = \zeta'_{B}$. 
From \eq.{eq:Utildeexpression}, we have 
\Equation{eq:Uplusminusexpression}
{19}{
     \mu^{*}_{(\pm)} 
       = \frac{(U^{(\alpha)}_{AA} \pm U^{(\alpha)}_{AB}) N(0)}
          {1 + (U^{(\alpha)}_{AA} \pm U^{(\alpha)}_{AB}) N(0) 
               \ln (W/\omega_{\rm c})} . 
     }
Therefore, we finally obtain 
\Equation{eq:AGeqforplusminus}
{20}{
     \ln \frac{T_{\rm c0}^{(\pm)}}{\Tc} = \Phi(\zeta^{(\pm)}) 
     =   \psi \bigl ( 
                      \frac{1}{2} + \frac{\zeta^{(\pm)}}{2\pi\Tc} 
              \bigr ) 
       - \psi \bigl (  \frac{1}{2}  \bigr ) 
     }
with $T_{{\rm c}0}^{(\pm)} = (2\e^{\gamma}/{\pi})
         \omega_{\rm c} \exp[-1/{\tilde \lambda}_{(\pm)}]$ and 
\Equation{eq:zetaplusminus}
{21}{
     \zeta^{(\pm)} = 
          (\frac{1}{2\tau_{1AA}}   - \frac{1}{2 \tau_{2AA}}) 
        + (\frac{1}{2\tau_{1AB}} \mp \frac{1}{2 \tau_{2AB}}) . 
     }

From \eq.{eq:AGeqforplusminus}, 
the isotope-effect coefficients for $\Tc^{(\pm)}$ are obtained as 
\Equation{eq:alpharesult}
{22}{
     \frac{\alpha^{(\pm)}}{\alpha_0^{(\pm)}}
     = \Bigl [  1  - 
          \frac{\zeta^{(\pm)}}{2\pi \Tc^{(\pm)} } 
          \psi'(\frac{1}{2} + \frac{\zeta^{(\pm)}}{2\pi\Tc^{(\pm)}})
       \Bigr ]^{-1} , 
     }
with 
${
     \alpha_0^{(\pm)} = \frac{1}{2} 
     [  1  -  
          ( {\mu^{*}_{(\pm)}}/{{\tilde \lambda}_{(\pm)}} 
          )^2 
     ] 
     }$. 
In general, $\alpha_0^{(\pm)}$ are largely different 
due to the difference in $\mu_{(\pm)}^{*}$. 
The physical origin of these differences is that 
the Coulomb energy $U_{AA}^{(\alpha)}$ is enhanced or reduced 
by $U_{AB}^{(\alpha)}$ depending on the signs $\pm$, 
as explicitly shown in \eq.{eq:Uplusminusexpression}.

For the $s$-wave pairing and the $s$-wave impurity scattering, 
$\zeta^{(+)} = 0$, while $\zeta^{(-)} = 1/\tau_{1AB}$. 
Thus, $\Tc$ of the state with $\Delta_{As} = \Delta_{Bs}$ 
is not suppressed by impurity doping at all, 
while that of the state with $\Delta_{As} = - \Delta_{Bs}$ 
is strongly suppressed, 
which is consistent with the results by 
Golubov and Mazin~\cite{Gol97} 
and Arseev {\it et al.}~\cite{Ars03} 
When $\Tc^{(+)} < \Tc^{(-)}$, 
the alternation of the solution and thus the jump of $\alpha$ occurs.

For anisotropic pairing and isotropic impurity scattering, 
since $1/2 \tau_{2XY} = 0$, we have $\zeta^{(+)} = \zeta^{(-)}$. 
Hence, the alternation of the solution does not occur.

\begin{figure}
\begin{center}
\includegraphics[width=5.5cm]{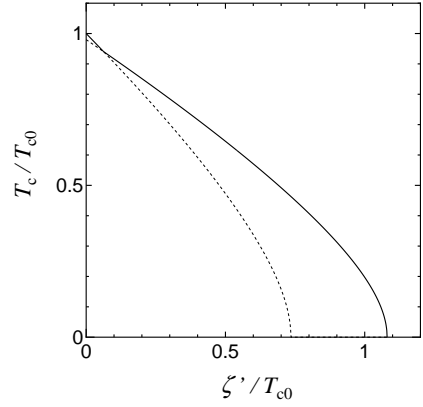}
\end{center}
\caption{
Impurity concentration dependence of the transition temperature $\Tc$. 
The parameter $\zeta'$ is proportional to the impurity concentration 
$n_{\rm imp}$. 
The solid line shows the resulting transition temperature 
for each $\zeta'$. 
The dotted line shows lower transition temperatures 
which are actually suppressed. 
} 
\label{fig:Tc}
\end{figure}

\begin{figure}
\begin{center}
\includegraphics[width=5.5cm]{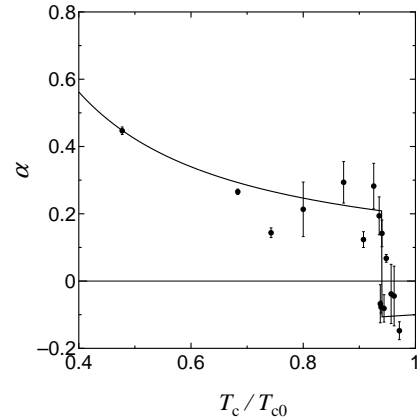}
\end{center}
\caption{
Isotope-effect coefficient as a function of transition temperature $\Tc$. 
The solid circles are the experimental data by Mao {\it et al.}~\cite{Mao01}
} 
\label{fig:alpha}
\end{figure}

Even for anisotropic pairing, it is possible that 
$\Lambda_{AB} = 1/2\tau_{2AB} \ne 0$, 
when $u_{AB}^{(2\alpha)} \ne 0$ or $v_{AB}^{(2\alpha)} \ne 0$. 
It is found in \eq.{eq:zetaplusminus} that 
when ${\tilde \lambda}_{AB} \Lambda_{AB} < 0$, 
one of $\Tc^{(\pm)}$ with lower $T_{\rm c0}$ decreases more gradually 
with impurity doping than the other. 
In this case, the gap function may alternate 
from $\Delta_{A\alpha} = \pm \Delta_{B\alpha}$ 
to $\Delta_{A\alpha} = \mp \Delta_{B\alpha}$ 
at some impurity concentration. 
In the transition from $\Delta_{A\alpha} = \pm \Delta_{B\alpha}$ 
to $\Delta_{A\alpha} = \mp \Delta_{B\alpha}$, 
the isotope-effect coefficient $\alpha$ jumps 
because of the difference in $\alpha_0^{(\pm)}$ mentioned above.

In Figs.~\ref{fig:Tc} and \ref{fig:alpha}, we show an example. 
To reproduce the experimental data of \SrRuO~\cite{Mao01}, 
we put $\alpha_0^{(+)} = - 0.1$, $\alpha_0^{(-)} = 0.2$, 
$\Tc_0^{(+)} = 1.5{\rm K}$, $\Tc_0^{(-)} = 0.98 \Tc_0^{(+)}$, 
and $\Lambda_{AB}/\zeta' = - 0.2$~\cite{Note03}. 
The values of $\Tc_0^{(\pm)}$ and $\alpha_0^{(\pm)}$ are 
reproduced, for example, 
by $\omega_{\rm c} = 410{\rm K}$~\cite{Mae97}, 
$\lambda_{AA} = 0.33678$, 
$\lambda_{AB} = 0.028509$, 
$\mu^{*}_{AA} = 0.16276$, 
and $\mu^{*}_{AB} = 0.028203$. 
In this case, since ${\tilde \lambda}_{AB} > 0$, 
we have $\Tc_0^{(+)} > \Tc_0^{(-)}$. 
In Fig.~\ref{fig:Tc}, the curves of $\Tc^{(\pm)}$ cross at 
$\zeta'/T_{{\rm c}0} \approx 0.063$ and $\Tc/T_{\rm c0} \approx 0.94$. 
Since the fold of the curve of $\Tc$ is slight for the present parameters, 
it would be hardly detected experimentally. 
At the crossing point, the superconducting state alternates 
from the state with $\Delta_{A\alpha} = \Delta_{B\alpha}$ 
to that with $\Delta_{A\alpha} = - \Delta_{B\alpha}$. 
In Fig.~\ref{fig:alpha}, it is shown that in this internal transition, 
$\alpha$ jumps. 
The present theory could reasonably reproduce 
the experimental result within the error bar of the data. 
In experiments, possible inhomogeneity of the samples might 
smear the transition.

The internal transition and the jump of $\alpha$ could occur 
in more general situations than in the above example. 
The eigenstates of the linearized gap equation, \eq.{eq:gapeqwithmatrix}, 
are subject to different impurity and Coulomb effects 
as shown 
in eqs.~(\ref{eq:Uplusminusexpression}) and (\ref{eq:alpharesult}) 
in the example. 
Since the eigenstate with the highest $\Tc$ occurs, 
the superconducting state alternates from one eigenstate 
to another with impurity doping for appropriate parameters. 
In the transition, the isotope-effect coefficient $\alpha$ jumps 
because the effective Coulomb parameter $\mu^{*}$ changes.

Lastly, we discuss the application to \SrRuO. 
The \SrRuO \hsp{0.25} compound has three electron bands, called 
$\alpha$, $\beta$, and $\gamma$, 
which have separate Fermi surfaces~\cite{Ogu95,Sin95}. 
There are some experimental results to support triplet pairing,~\cite{Ish98} 
but the momentum dependence of the gap function 
is controversial~\cite{Tan01,Iza01,Mae00}. 
As argued above, the present mechanism holds, 
as long as the parameters satisfy the condition for the alternation 
of the eigenstates, 
independently of the number of electron bands and 
the momentum dependence of the gap function, 
whether additional nonphonon pairing interactions exist or not~\cite{Note02}. 
However, for the alternation at $\Tc \approx 0.94 \Tc_0$ 
to be quantitatively reproduced, 
it is necessary that two of the eigenstates have very close $\Tc_0$. 
It could occur as a result of a combination of 
three electron bands. 
The split of the critical field curve observed 
for parallel fields~\cite{Deg02} 
may suggest an existence of the hiden eigenstate 
slightly below $\Tc_0$ at the zero field.

We may consider another possibility for \SrRuO. 
The internal degrees of freedom play an essential role 
in the present mechanism. 
In the present model, they originate from the multiband nature, 
but they may originate from the anisotropic gap structure 
of triplet pairing superconductivity. 
If we apply the above calculation to a model with a pair of the 
eigenstates with slightly different $\Tc_0$'s 
by replacing the band suffixes $A$ and $B$ 
with the suffix $\alpha$ to express the momentum dependence, 
the same figures as Figs.~\ref{fig:Tc} and \ref{fig:alpha} 
are reproduced for appropriate parameter values. 
The explicit calculation will be presented in a separate paper.

In conclusion, we have examined the impurity effect in multiband 
superconductors, and obtained the expressions of $\Tc$ and $\alpha$. 
It has been found that an internal transition of the superconducting 
state is induced by impurity doping in some condition, 
and the transition is accompanied by a jump of 
the isotope-effect coefficient $\alpha$. 
The $\Tc$ dependence of $\alpha$ deviates from the CGP universal 
relation due to the jump. 
It is possible under appropriate parameters 
that $\Tc$ appears to obey the standard AG equation 
simultaneously with a large deviation from the CGP relation.

The author wishes to thank Y.~Maeno and Z.Q.~Mao for useful 
discussions and for sending their experimental data to him.



\end{document}